\documentclass[11pt,a4paper,onecolumn]{IEEEtran}

\usepackage{amsmath,amsthm}
\usepackage{graphicx}
\usepackage{float}
\usepackage{setspace}
\usepackage{amsthm}

\usepackage{cite}
\usepackage{subfig}
\graphicspath{{plots/}}
\usepackage{todonotes}
\newtheorem{thm}{Theorem}
\newtheorem{prop}{Proposition}

\def\Rnet{\tilde{{\cal R}}_0}
\def\R{{\cal R}_0}
\def\tR{\tilde{\cal R}_0}
\begin{document}
%
% paper title
% can use linebreaks \\ within to get better formatting as desired
\title{Poisson Network SIR Epidemic Model}

% author names and affiliations
% use a multiple column layout for up to two different
% affiliations

%}

% conference papers do not typically use \thanks and this command
% is locked out in conference mode. If really needed, such as for
% the acknowledgment of grants, issue a \IEEEoverridecommandlockouts
% after \documentclass

% for three or more authors, or if they all won't fit within the width
% of the page, use this alternative format:
%
\author{
    \IEEEauthorblockN{Josephine K. Wairimu\IEEEauthorrefmark{1}\IEEEauthorrefmark{2}, Andrew Gothard\IEEEauthorrefmark{2}, and 
    Grzegorz A. Rempala\IEEEauthorrefmark{2}
    }
    \\
    \IEEEauthorblockA{\IEEEauthorrefmark{1}University of Nairobi, Kenya: Dept. of Mathematics\\
    Nairobi, Kenya\\
    Email: kagunda.2@osu.edu}
    \IEEEauthorblockA{\IEEEauthorrefmark{2}The Ohio State University: Division of Biostatistics\\
    Columbus, 43210, USA\\
    Email: rempala.3@osu.edu}
}
%\IEEEauthorblockA{\IEEEauthorrefmark{3}line 1 (of Affiliation): dept. name of organization\\
%line 2: name of organization, acronyms acceptable, City, Country\\
%line 3: Email: name@xyz3.com}
%}

% use for special paper notices
%\IEEEspecialpapernotice{(Invited Paper)}

% make the title area
\maketitle

\begin{abstract}
We extend the classical Susceptible-Infected-Recovered (SIR) model to a network-based framework where the  degree distribution of nodes follows a Poisson distribution. This extension incorporates an additional parameter representing the mean node degree, allowing for the inclusion of heterogeneity in contact patterns. Using this enhanced model, we analyze epidemic data from the 2018-20 Ebola outbreak in the Democratic Republic of the Congo, employing a survival approach combined with the Hamiltonian Monte Carlo method. Our results suggest that network-based models can more effectively capture the heterogeneity of epidemic dynamics compared to traditional compartmental models, without introducing unduly overcomplicated  compartmental framework.
\end{abstract}


\newcommand{\JK}[1]{\todo[color=cyan]{\small{#1 --- JK}}} % Josephine 
\makeatother
%\doublespacing
%%%%%%%%%%%%%%%%%%%%%%%%%%%%%%%%%%%%%

\section{Introduction:SIR Compartmental Modeling}\label{sec:1}

The study of the causes, transmission, and control of infectious diseases has a rich history, dating back to Grant's work in 1620 \cite{FredBrauer}. In mathematical epidemiology, mathematical models and analytical techniques are employed to understand the spread and control of infectious diseases. These models enable researchers to predict disease dynamics, assess the impact of various interventions, and guide public health strategies.

One of the earliest examples of such modeling comes from Daniel Bernoulli \cite{Bernoulli1760}, who in 1760 applied mathematical principles to study the trade-offs between the benefits and risks of variolation during a deadly smallpox outbreak in the American colonies. His approach classified individuals based on their epidemiological status, laying the groundwork for what are now known as compartmental models.

The first malaria compartmental model was developed by Sir Ronald Ross in early 1900's. He used human and vector compartments to demonstrate that malaria was transmitted by mosquitoes to humans \cite{ROSS}. In 1927, the classical Susceptible-Infected-Recovered (SIR) compartmental model was introduced by Kermack and McKendrick to describe the transmission of infectious diseases \cite{KMK1927, KMK1932, KMK1933}. For a comprehensive review of both classical and more recent SIR-type models, as well as related developments, see \cite{FredBrauer} and the references therein.

In the original Kermack-McKendrick model, individuals in a population are divided into three compartments: susceptibles ($S$) infected ($I$) and recovered ($R$), with the proportion of   $S,I,R$ type  individuals at time $t$  denoted by $S(t)$,  $I(t)$, and  $R(t)$ respectively. The system of ordinary differential equations (ODEs) governing the transfer of individuals between these compartments is given by

\begin{equation}
\begin{aligned}
\dot S &= -b\, S\,I\\
\dot I &= b\, S\,I- c\, I \\
\dot R &= c\, I
\end{aligned}
\label{eq:sir}
\end{equation}
where dotted quantities denote time derivatives and where for simplicity we suppressed the time argument in functions $S, I,$ and $R$.  

The above system is known as the SIR compartmental ODE model, the simplest example of a deterministic system describing the spread of a disease in a closed population. From the equations, we note the following.
\begin{itemize}
\item $b$ represents the rate at which susceptible individuals become infected upon encountering infected individuals (the infection rate). 
\item $c$ defines the rate at which infected individuals recover or succumb, thus leaving the infected compartment (the recovery rate).
\item $\R = b/c$ is a threshold parameter known as the {\em basic reproduction number} which  represents the average growth rate of secondary cases in a completely susceptible population  \cite{FredBrauer}.
\item  When $\R > 1$, the epidemic will, at least initially, spread within the population as the number of infected individuals increases. In contrast, if
$\R \leq 1$, the number of infected individuals will decline and the epidemic will not spread.
  
\end{itemize}

If $0 < d \ll 1$ represents a small proportion of initially infected individuals, the initial conditions of the system are given by $(S(0), I(0), R(0)) = (1, d , 0)$. From the form of \eqref{eq:sir}, it follows that the system satisfies the conservation law $S + I + R = 1+d$. This and elementary manipulations of the first and the last equation in \eqref{eq:sir}  give the following alternative representation of the classical SIR model 

\begin{equation}
\begin{aligned}
\dot S &= -b\, S\,(1+d-S+\R^{-1}\log S)\\
\dot I &= -\dot S- c\, I \\
R &= 1+d-I-S.
\end{aligned}
\label{eq:sir2}
\end{equation}
Note that the first equation in the system above involves only \( S \). Since it describes the decline of susceptibles (and consequently the emergence of new cases), it is often referred to as the {\em epidemic curve equation} (see, for instance, \cite{RempalaBiomath}). 

For the SIR model to effectively describe the disease dynamics in a given population of interest, it is typically necessary to estimate the parameters \( b \), \( c \), and \( d \).

While compartmental ODE models are valuable tools in epidemiology, they have several notable limitations \cite{RempalaDSA2023}. First, these models rely on the law of mass action, which assumes uniform social contact patterns across the population. This simplification neglects the inherent heterogeneity in individual behaviors and fails to account for changes in contact patterns as an epidemic unfolds, a crucial factor for diseases where person-to-person transmission is the primary mode of spread. 

Additionally, as deterministic models, they overlook the dynamics of early outbreak phases, during which the number of infections is small and stochastic events play a significant role in shaping disease transmission. As disease dynamics grow more complex—such as when multiple types of infected or susceptible individuals are involved—these models become increasingly intricate, complicating interpretation and analysis.

Given these limitations, a different approach, often based on stochastic modeling, is frequently more effective \cite{RempalaDSA2023,DiLauro2022}. While stochastic models are typically more challenging to formulate, they better accommodate heterogeneity and can simplify the overall structure by reducing the number of compartments. Furthermore, they inherently capture variability by incorporating random fluctuations, offering a more realistic representation of disease spread in diverse populations \cite{schwartz2015estimating}. 

This paper explores the classical SIR model from a stochastic perspective, with a focus on its extension  to the network-based SIR model, where transmission dynamics are governed by a static contact network. Specifically, we highlight the lesser-known relationships between the classical SIR model, its Poisson network counterpart, and the pairwise closure condition. Through a real data example, we demonstrate that while both models have similar computational complexity, the network-based model holds the advantage of explicitly accounting for and revealing the network's degree distribution.

The rest of the paper is organized as follows: In Section~\ref{sec:2}, we provide background on network-based stochastic SIR models, focusing on configuration models, their pairwise representations, and dynamical survival approaches to model fitting. Following \cite{RempalaBiomath}, we argue that in some cases the classical SIR model serves as a good approximation of the Poisson SIR network model. 
In Section~\ref{sec:3}, we address the statistical inference problem for stochastic network-based SIR models under a Poisson degree distribution and derive the relevant likelihood equations for parameter estimation. We then analyze the 2018-20 Ebola  outbreak data in the Democratic Republic of the Congo, fitting the data to the derived network model and comparing the result to similar analysis based on the classical SIR approach presented in \cite{Vossler}.
Finally, in Section~\ref{sec:4}, we discuss the  results and present conclusions from our work.

\section{Stochastic Network SIR  Model}\label{sec:2}

Stochastic network models address some of the challenges discussed earlier by providing a more realistic framework for capturing and studying the complexities of disease spread in structured populations. These models account for the variability in individual interactions and transmission dynamics.
A stochastic model represents contact patterns during an epidemic as a graph, where nodes correspond to individuals, and edges denote potential transmission routes. The stochastic SIR epidemic process on a network of size \( n \) can be described as follows.

At the onset of the epidemic, \( m \) individuals are randomly selected as initially infectious. Each infectious individual remains in this state for a duration drawn from an exponential distribution with rate \( \gamma \). During this period, the individual contacts their immediate neighbors according to a Poisson process with intensity \( \beta \). If a contacted neighbor is susceptible, they become infectious immediately. Once the infectious period ends, the individual recovers and becomes immune to further infections. All infectious periods and Poisson processes are assumed to be independent (see, e.g., \cite{Gray2011}).

\begin{figure}
    \centering
    \includegraphics[width=0.4\linewidth]{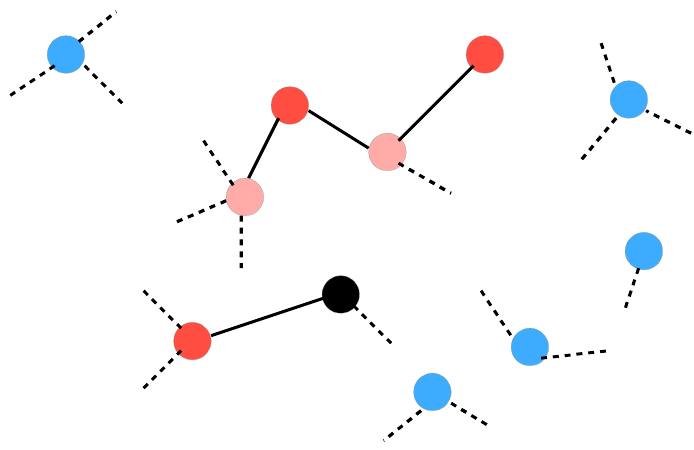}
    \caption{{\bf SIR Dynamics on Network.} Blue nodes represent susceptible individuals, while  red and pink ones represent  the initially  infected and secondarily infected individuals, respectively. The black node indicates a removed individual.  Dashed half-edges connect uniformly at random to form solid edges.}
    \label{fig:CMnet}
\end{figure}

In many applications, it is useful to track the spread of an epidemic while simultaneously constructing the underlying transmission network. This can be achieved by generating a random graph in tandem with modeling the epidemic's spread. The process, illustrated in Figure~\ref{fig:CMnet}, begins by assigning to each node a number of unconnected half-edges based on the degree distribution \( p \). Static connections are then formed by matching the half-edges of infectious nodes to other available half-edges in the network as part of the Poisson contact process with intensity \( \beta \), as described earlier. If the connected node is susceptible, transmission occurs. Otherwise, the infectious individual attempts to connect its remaining half-edges to other available half-edges, repeating this process until all half-edges are matched or no further connections are possible. While this matching process may occasionally result in self-loops or multiple edges between two nodes, such occurrences are negligible in the limit of large \( n \).

The network SIR process as described above  is seen to be a continuous time Markov chain \cite{Andersson-Britton} on a configuration model random graph $\mathcal G(n,p)$ of size $n$ and the degree distribution $p=(p_k)_{k=0}^\infty$.  Recall that the probability generating function (PGF) corresponding to $p$ is defined as 
$$\psi(x)=\sum_{k=0}^{\infty}p_kx^k.$$ 
The average  degree is then defined in terms of PGF $\psi$  as $\mu=\psi^{\prime}(1)$. Similarly,  the average excess degree (i.e., the  average degree of a random neighbor of a node, minus one) is defined as $\mu_{ex} = \psi^{\prime\prime}(1)/\mu$ (see, e.g., \cite{Newman2002,RempalaBiomath}).
The probability generating function is needed here as it tells us about the properties of a randomly chosen susceptible node. We use the PGF to find the probability that a randomly chosen node $u$, that is initially susceptible, in a infinite network, remains so up to time $t$.

%$$$$$$$$$$

\subsection{Pairwise Model and Its Closure}

The pairwise model provides a comprehensive way of describing the dynamics of an SIR epidemic on a configuration model graph. The pairwise model equations (see, for instance, \cite{Kiss2023}) are:

\begin{equation}\label{eq:pwmod}
\begin{aligned}
\dot{[S]} & = -\beta[SI]\\
\dot{[I]} & = \beta[SI] - \gamma[I]\\
\dot{[R]} & = \gamma[I] \\
\dot{[SI]} & = -\gamma[SI] + \beta([SSI] - [ISI]) - \beta[SI]\\
\dot{[SS]} & = -2\beta[SSI]
\end{aligned}
\end{equation}

% counting integer values
 where $[A]$, $[AB]$, $[ABC]$ with $A, B, C \in \{S, I, R\}$ stand for the functions that at their integer values may be interpreted as counts of the number of singles, doubles and triples in the entire network with the given sequence of states when each group is counted in all possible ways.
More formally, at integer values 
\begin{equation*}
    [ABC] = \sum_{i = 1}^N \sum_{j = 1}^{N} \sum_{k=1}^N a_{ij} a_{jk} I_i(A) I_j(B) I_k(C)
    \label{eq:ABC}
\end{equation*}
where $(a_{ij})_{i,j=1,2,\dots, N}$ is the adjacency matrix of the network with entries either zero or one and $I_i(A)$, $I_i(B)$, and $I_i(C)$ are binary variables that equal one when the status of $i$-th individual is $A$, $B$, and $C$, respectively, and equal zero otherwise. The singles $[A]$ and doubles $[AB]$ are similarly defined.

The model described in \eqref{eq:pwmod} is not especially useful as stated,  since clearly additional equations for triples dynamics are required, which, in turn, depend on dynamics of quadruples, and so on, leading to an escalating number of variables and equations. To manage this complexity and make the model tractable, a "closure" approach is often employed. In this approach, larger structures (e.g., triples) are approximated in terms of smaller ones (e.g., pairs). Indeed, The model described in \eqref{eq:pwmod} can be closed at the level of pairs by applying the following result recently presented in  \cite{Kiss2023}.  

\begin{thm}\label{thm:1}[Exact pairwise closure] Consider the SIR Markov process  on the configuration graph $\mathcal G(n,p)$ as described above. Let $\kappa= \mu_{ex}/\mu$. The system \eqref{eq:pwmod} may be closed exactly by setting 
\begin{equation*}\label{eq:cls2}
    [ASI]\simeq\kappa \frac{[AS][SI]}{[S]}    
\end{equation*}
for $A \in \{S, I\}$ iff the  degree distribution $p$  is  binomial, Poisson  or negative binomial.  The closure  is exact in the sense that the equality in \eqref{eq:cls2} with both sides  multiplied by $n^{-1}$ holds asymptotically as $n\to \infty$. 
Furthermore, $\kappa = (k - 1) / k < 1$, $\kappa = 1$, or $\kappa = (r + 1) / r > 1$ if the degree distribution is binomial $\mathrm{BINOM}(k,q)$, Poisson $\mathrm{POI}(\lambda)$, or negative binomial $\mathrm{NB}(r,q)$, respectively.
\end{thm}

We direct interested readers to the original paper for a formal statement of this result, along with its proof and additional details that are beyond the scope of this discussion.

Applying  the closure relation in Theorem~\ref{thm:1},  the system  (\ref{eq:pwmod}) may be written as  
\begin{equation}
\begin{aligned}
\dot{[S]} & = -\beta[SI]\\
\dot{[I]} & = \beta[SI] - \gamma[I]\\
\dot{[R]} & = \gamma[I] \\
\dot{[SI]} & = -\gamma[SI] + \beta\kappa\frac{[SI]([SS]-[SI])}{[S]} - \beta[SI]\\
\dot{[SS]} & = -2\beta\kappa\frac{[SI][SS]}{[S]}.
\end{aligned}
\label{eq:pwmod2}
\end{equation}
% we can figure out what [SS] is.
The above system may be considerably simplified, by removing last two equations. Indeed, by dividing the last equation by the first one and solving the resulting differential equation under the assumption that $[SS](0)=\mu[S](0)$, it follows that $[SS]=\mu[S]^{2\kappa}$.  Substituting now this formula into the fourth equation  and dividing again by the first one we obtain another differential equation for $[SI]$ as a function of $[S]$. That  equation may be then solved explicitly, depending on the value of $\kappa$,  yielding the reduced version of \eqref{eq:pwmod2} which by Theorem~\ref{thm:1} may be written in terms of the limiting proportions of $[S]/n$, $[I]/n$, and $[R]/n$ as $n\to \infty$  which we denote  below by  $S, I, R$.  The final system has the following form:  
%\begin{equation}
\begin{align}\label{eq:dsa1}
    -\dot{S} &= \nonumber\\ 
    &\begin{cases}
        \tilde{\beta}(S^{\kappa}-S^{2\kappa})+\frac{\tilde{\gamma}}{1-\kappa}S(1-S^{\kappa-1})+\tilde{\rho} S^{\kappa} 
        &\text{if } \kappa\ne 1 \\
        \tilde{\beta} (S - S^{2}) + \tilde{\gamma} S \log S+ \tilde{\rho} S
        &\text{if } \kappa = 1 
    \end{cases} \nonumber\\
    \dot{I} &= -\dot{S} -\gamma I\\
    R & = 1+\rho-S-I, \nonumber  
\end{align}
%\end{equation}
where the initial conditions are (per analogy with   the classical SIR model \eqref{eq:sir} in Section~\ref{sec:1}) $S(0)=1, I(0)=\rho, R(0)=0$  and  $\tilde{\rho} = \beta \mu \rho$,  $\tilde{\gamma}=\beta+\gamma$, and $\tilde{\beta}=\mu \beta$.

Note that the equations above may be interpreted as the mean field approximation to  the scaled SIR  Markov process evolving on the configuration model random graph ${\cal G}(n,p)$ as described in the beginning of this Section. 

\subsection{Poisson Network SIR}

For the reminder of the paper we will consider only the special case of the Poisson network SIR model, namely when  $p$ distribution in ${\cal G}(n,p)$ is of the form $p_k =\exp{(-\mu)}\mu^k/{k!}$ for $k\ge 0$. 
This gives the PGF formula 
$$\Psi(x)=\exp{\{\mu(x-1)\}}  $$
and consequently implies that $\mu=\Psi^\prime(1)= \Psi^{\prime\prime}(1)/\mu=\mu_{ex}$, so that $\kappa=1$ (see also Theorem~\ref{thm:1}). 
In this case the system~\eqref{eq:dsa1} takes the form 
\begin{align} \label{eq:dsa2}
    \dot{S} &= 
     -\tilde{\beta}\, S(1+\rho - S + \tR^{-1}  \log S) \nonumber\\
    \dot{I} &= -\dot{S} -\gamma I\\
    R & = 1+\rho-S-I, \nonumber  
\end{align}
where $\tR =\tilde{\beta}/\tilde{\gamma}$ is a network analogue of the basic reproduction number discussed in Section~\ref{sec:1}.  In general,  \eqref{eq:dsa2} is seen to be  algebraically closely related to \eqref{eq:sir2} as formalized in the following result (see also \cite{RempalaBiomath}).

\begin{figure}[htb]
    \centering
\includegraphics[width=.6\linewidth]{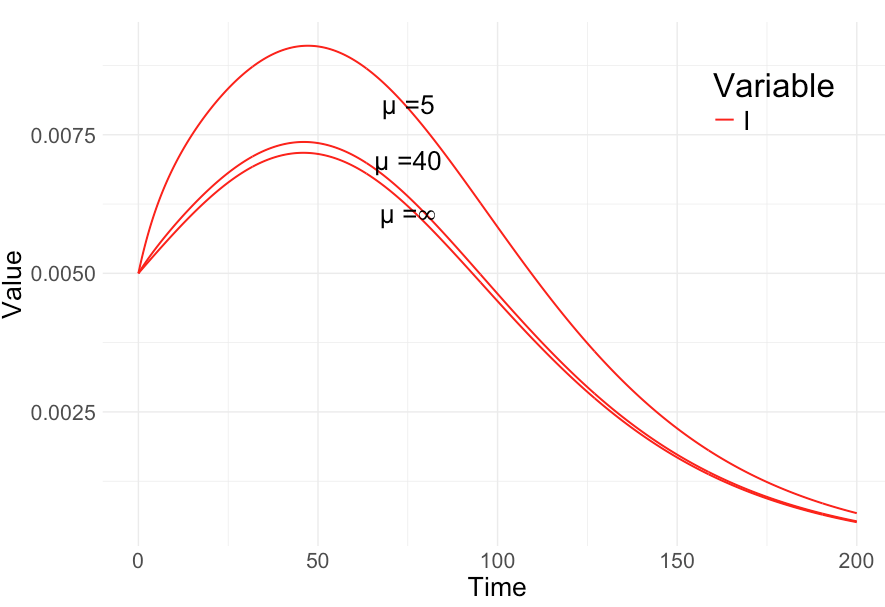}
    \caption{{\bf Approximating Network SIR Model}.  A simple example illustrating the result of  Proposition~1. For large mean degree distribution $\mu$ (here at least 40), the network curves for infected are seen to get close to the one corresponding to the classical SIR model (lowest curve). For this particular example the  values of the parameters $(\tilde\beta,\tilde\gamma,\rho)$  are taken  from the first column of Table~\ref{tab:1} in Section~\ref{sec:3} below.}
    \label{fig:2}
\end{figure}

\begin{prop}\label{prop:1}
Assume that we wish to approximate the dynamics of a Poisson network SIR epidemic given by \eqref{eq:dsa2} using the classical SIR equations \eqref{eq:sir2}. The two models' respective epidemic curve equations (and thus also respective $S$-curves) coincide iff $b =\tilde\beta$, $d =\rho$
and $c = \tilde\gamma$. In this case $\R=\tR$. Moreover, if $\R <\mu$ than the 
true  infection curve under the network model  satisfies
$$\dot{I}=-\dot{S} -c\,(1-\frac{\mathcal R_0}{\mu})\, I.$$ 
\qed\end{prop}
The result above indicates that the Poisson network SIR model will closely approximate the classical SIR model in terms of its Susceptible-Infected-Recovered (SIR) curves when the mean degree of the Poisson network ($\mu$) is high. In this scenario, the  factor $1-{\mathcal R_0}/{\mu}$approaches one. As the mean degree increases, the network model converges to the classical SIR model because the high connectivity reduces the impact of individual-specific interaction patterns. Consequently, the infection and recovery dynamics become more uniform, aligning closely with the simpler SIR framework. This is further illustrated in Figure~\ref{fig:2} where the the $I$-curves of the Poisson network SIR model are plotted (in red) for several different values of $\mu$. The lowest curve corresponds to $\mu=\infty$ which is the classical  SIR model represented by ODEs \eqref{eq:sir} and \eqref{eq:sir2}. 

The system \eqref{eq:dsa2}, for numerical convenience and to avoid evaluating terms of the form $\log S$ when $S$ is small, is often reformulated into an alternative representation of the Poisson SIR equation. Specifically, by introducing the substitution  
$$ 
D = 1 + \rho - S + \tR^{-1}\log S, 
$$  
and differentiating, we can reformulate the first equation in terms of two others as shown in \eqref{eq:dsa3}. These two equations resemble the classical SIR equations \eqref{eq:sir}, with $I$ replaced by $D$, and with the parameters matched as described in Proposition~\ref{prop:1}. 
\begin{align} \label{eq:dsa3}
    \dot{S} &= 
     -\tilde{\beta}\, D S \nonumber\\
     \dot{D} & = -\dot S -\tilde \gamma D \nonumber\\
    \dot{I} &= -\dot{S} -\gamma I\\
    R & = 1+\rho-S-I, \nonumber  
\end{align}
with additional initial condition $D(0)=\rho$.

As  noted in  \cite[Chapter~2]{keeling2008modeling} for better numerical stability, it is sometimes convenient to rewrite the above set of equations in terms of log transformed quantities $S_\ast=\log S$, $I_\ast =\log I$, and $D_\ast = \log D$ as 
\begin{align} \label{eq:logdsa}
    \dot{S}_\ast &= 
     -\tilde{\beta}\, \exp(D_\ast)  \nonumber\\
     \dot{D}_\ast & = \tilde\beta\, \exp(S_\ast) -\tilde \gamma \\
    \dot{I}_\ast &= \tilde\beta\exp(S_\ast+D_\ast-I_\ast) -\gamma,  \nonumber
\end{align}
with $R$ recovered via the conservation law as before. 
\subsection{Dynamical Survival Analysis}

Although \eqref{eq:dsa2} is derived from the mean-field approximation of a Markov process evolving on an infinite random Poisson degree graph, it also serves as a basis for approximate inference regarding both the parameters of the underlying network and the SIR process. This method, which employs ODEs to characterize a stochastic process and is referred to as {\em dynamical survival analysis} or DSA in \cite{DiLauro2022}, was originally proposed in \cite{Khud-SDS}. In that work, it was argued that, for a randomly selected node, the $S$ function described in \eqref{eq:dsa2} can be interpreted as an improper survival function (see also \cite{RempalaDSA2023}). Specifically, if 
$T_I$ denotes the random time at which a randomly chosen, initially susceptible  node in the network becomes infected, then in the limit of the large graph (that is, as the number of nodes $n\to \infty$) we have for any $t>0$ 
$$P(T_I>t) = S(t).$$
Note that taking in \eqref{eq:dsa2} $\dot S(\infty) =0$ (since at the end of the epidemic there is no further depletion of susceptibles, see e.g, 
\cite{RempalaDSA2023}) gives  $S(\infty)=1-\tau$
where \begin{equation}\label{eq:tau}   \tau = \tR^{-1}\log(1-\tau)-\rho
\end{equation}is the final epidemic size or, equivalently, the limiting probability of a randomly chosen initially susceptible node being ever infected, i.e., $\tau=P(T_I<\infty)$.

More generally, if  for $T>0$ we define \begin{equation}\label{eq:tauT}
    \tau_T =1 - S(T)
\end{equation} then 
 the quantity \begin{equation}\label{eq:f}
    f_{\tau_T}(t)=-\dot S(t)/\tau_T
\end{equation} may be considered as a  density function for the conditional random variable $T_I^\ast=T_I|(T_I<T)$. Notice that we may take here $T=\infty$ in which case $ \tau_\infty =\tau$, the quantity defined in \eqref{eq:tau}. In view of this, since we may  use the second equation in  \eqref{eq:dsa2} to represent the variable $I$  in terms of a convolution: 
\begin{equation*}
    I(t)=\rho\exp(- \gamma\, t) +\int_0^t (-\dot S(u))\exp\{-\gamma(t-u)\}du, 
    \end{equation*}
    we see that the quantity 
    \begin{equation}\label{eq:g}
      g_{\tau_T}(t) = \frac{\gamma}{\tau_T}(I(t)-\rho\exp(-\gamma t))  
    \end{equation} is simply a density function of a random variable, say $T_R$, that is a sum of two independent random variables: $T_I^\ast$ and $U$ where $U$ is an exponential variable with rate $\gamma$. 
    It follows that the system \eqref{eq:dsa2} describes the densities of a pair of  random variables $(T_I^\ast,T_R)$ representing the history of epidemic of a randomly chosen network node in an infinite network.  By construction, these variables are  such that $T_I^\ast$ is independent of the exponential variable $U=T_R-T_I^\ast$ which is interpreted as the infectious period.

Note that the vector of parameters describing the distribution of the pair \((T_I^\ast, T_R)\) is 
\begin{equation}\label{eq:theta} 
\theta = (\tilde \beta, \tilde \gamma, \gamma, \rho)
\end{equation} 
as all other relevant quantities, such as \(\tau_T\), \(\mu\), and \(\tR\), can be expressed in terms of \(\theta\) or \(S\), which is fully determined by \(\theta\). 
    Indeed,  recall  that $\tR = \tilde\beta/\tilde\gamma$, $\tau_T= 1-S(T)$ with $\tau_\infty$  defined by \eqref{eq:tau} and that for the  average degree $\mu$ we have 
    \begin{equation}\label{eq:invmu}
     \mu = \tR\left(1-\frac{\gamma}{\tilde\gamma}\right)^{-1}.    
    \end{equation}
In view of the above discussion,   it is sometimes more convenient to consider an alternative parametrization  

\begin{equation}\label{eq:theta_pr} 
\theta^\ast = (\tilde \beta, \tilde \gamma, \mu, \rho), \end{equation} which is the parametrization used in our numerical example below.

\section{Statistical Inference}\label{sec:3}

\subsection{Ebola Dataset}
We will illustrate the practical application of the network-based epidemic model and the approximation result in Proposition~\ref{prop:1} by analyzing the  data from the third and final wave of the 
2018–2020 Ebola Virus Disease (EVD) outbreak in the Democratic Republic of the Congo (DRC). The complete dataset from this 
outbreak is described and analyzed in detail in \cite{Vossler}; here, we focus specifically on a subset of the data, which includes information on the onset and 
hospitalization of 1,069 confirmed EVD patients recorded between May 27, 2019, and September 12, 2019. This subset is 
derived from the broader EVD database, which is 
comprehensively detailed in Section~2 of \cite{Vossler}.

Since early August 2018, the DRC Ministry of Health, in collaboration with 
several international partners, has been working to support and strengthen the response to the Ebola Virus Disease (EVD) 
outbreak. This effort has been coordinated through the 
Emergency Operations Center in Goma, the capital of North Kivu 
Province in eastern DRC. Despite regional security challenges~\cite{kraemer2020dynamics}, response teams were 
deployed wherever possible to interview patients and suspected contacts. These interviews were conducted using a standardized 
case investigation form. Based on the collected information, cases were categorized as suspected, probable, or confirmed.

A suspected case (whether the patient survived or not) was defined as one with an acute onset of fever (over 100$^{\circ}$F) and at least three Ebola-compatible clinical 
signs or symptoms, including headache, vomiting, anorexia, diarrhea, lethargy, stomach pain, muscle or joint aches, difficulty swallowing or breathing, hiccups, unexplained 
bleeding, or any sudden, unexplained death. If a patient met the suspected case definition but died, and no specimens were available for testing, the case was considered probable. A 
confirmed case of EVD was defined as a suspected case with at least one positive test for Ebola virus using reverse transcription polymerase chain reaction (RT-PCR)~\cite{maganga2014ebola}.

Patients suspected of having EVD were isolated and transported to an Ebola Treatment Center (ETC) for 
confirmatory testing and treatment~\cite{ilunga2019ongoing}. Following the approach in \cite{Vossler}, our analysis of the DRC dataset focused on the dates of symptom onset and removal, 
with "removal" defined as either death or recovery at home, or transfer to an ETC. It was assumed that, once patients were admitted to an ETC, the probability of further infection 
spread was minimal due to strict safety protocols. This assumption was later reinforced by the vaccination of 
healthcare personnel and family members who had been in contact with the suspected Ebola cases.

\subsection{Statistical Model}
The probabilistic DSA model used to derive \eqref{eq:f} and \eqref{eq:g} can now be applied in a similar manner to derive the {\em likelihood function} for  inference of the parameters vector $\theta$ given in \eqref{eq:theta}. The justification for this approach is discussed in \cite{Khud-SDS} through the so-called Sellke construction. We omit the details here, referring interested readers to the original source. Instead, our focus will be on the mechanics of the statistical inference process.

\paragraph{Likelihood} We assume that observations of new infections and 
recoveries are available up to some time horizon $T$ such that  $T \in [0,\infty]$. Recall the definition of \eqref{eq:tauT} and 
note that  $\tau_T$ is non-decreasing in $T$ and that $\tau_\infty= \tau<1$, the final 
epidemic size given by \eqref{eq:tau}. As already indicated, the density \eqref{eq:f} 
may be interpreted as a conditional density function of infection times $T_I^\ast$ on $[0,T]$ for any $T\le\infty$. Recalling the 
pair of variables $(T_I^\ast,  T_R) $  and the fact that the variable $W=T_R-T_I^\ast$  
is exponentially distributed with rate parameter $\gamma$  we obtain the following  
individual-level likelihood function for an initially susceptible individual $i$ 
observed until time $T$ with infection and recovery times $t_i$ and $r_i$, respectively:   
\begin{equation}\label{eq:lkh1}
\mathcal{L}(\theta\vert t_i,r_i,T)=f_{\tau_T}(t_i)\,\gamma^{w_i}e^{-\gamma (r_i \wedge T -t_i)}.\end{equation}

Here, $w_i$ is the event indicator, satisfying $w_i = 0$ if $r_i \wedge T = T$ and $w_i = 1$ otherwise.  
The likelihood for the set of $n$ individuals in the population with complete records \eqref{eq:lkh1} is  the product of the
individual likelihoods, reflecting the assumption that the infection events  are (approximately) independent in a large population.   

\paragraph{Likelihood with missing data} As discussed in \cite{Vossler} in 
about 30\% of the DRC Ebola cases the individual disease records were incomplete, missing either  infection $(t_i)$ or 
recovery $(r_i)$ times.  Fortunately,  such missingness may be handled by  the DSA likelihood   without any  need for data 
imputation. In case when  only $t_i$  is observed,  ($r_i$   is missing),   the 
likelihood \eqref{eq:lkh1} reduces  simply to \eqref{eq:f} \begin{equation}\label{eq:lkh2}
\mathcal{L}(\theta\vert t_i,\circ,T)=f_{\tau_T}(t_i).\end{equation}
On the other hand, if only $r_i$ is observed ($t_i$ is  missing),   the  likelihood is obtained from  the convolution formula and \eqref{eq:g}
\begin{equation}\label{eq:lkh3}  \mathcal{L}(\theta\vert \circ,r_i,T) = g_{\tau_T}(r_i).\end{equation} 
Similarly as above, the likelihood for incomplete data \eqref{eq:lkh_inc} is obtained by taking product of \eqref{eq:lkh2}
and \eqref{eq:lkh3} over all individual incomplete histories (see also next subsection). 

\paragraph{Effective  population size and outbreak size}

In many outbreak datasets, including the one from DRC considered here,  only infection and recovery times  are recorded. It  is therefore often difficult to determine the size ($N$) of the susceptible population at risk of infection.  This is known in the literature as the problem of estimating  the {\em effective population size}~\cite{Khud-SDS}. Under the DSA  model, this estimate may be obtained as 
\begin{equation}
  \hat{N}=\frac{k_T}{\tau_T},
  \label{eq:nhat}
\end{equation} 
where $k_T$ is the count of observed infected over the time horizon $T$ and $\tau_T$ is given by \eqref{eq:tauT}.  Similarly, one may also estimate the final epidemic count $K_\infty$ of all already observed  and future infections  by 
\begin{equation}
  \label{eq:khat}
  \hat{K}_\infty = \hat{N} \tau,
\end{equation} 
where  $\tau = \lim_{T\to \infty} \tau_T $ is the final epidemic size.

\subsection{Parameter Estimation}\label{ssec:par_est}

As discussed in the previous section, the epidemic history of the \(i\)-th individual can be represented either by the times of disease onset and removal \((t_i, r_i)\) or by incomplete 
data, such as \(t_i\) or \(r_i\) alone, denoted as \((t_i, \circ)\) or \((\circ, r_i)\) (\(\circ\) indicates a missing value). Among the available \(n_T\) individual recorded  histories, we assume there are \(n\) complete records \((t_i, r_i)\), \(n_1\) incomplete records of type \((t_i, \circ)\), and \(n_2\) incomplete records of type \((\circ, r_i)\). 

The  DSA likelihood function for the \(n\) complete data records is derived from \eqref{eq:lkh1} and takes the form:
\begin{equation}\label{eq:lkh_comp}
	\begin{aligned}
		&\mathcal{L}_C(\theta\vert t_1,\ldots,t_n, r_1,\ldots,r_n, T) \\ 
		&= \prod_{i=1}^n f_{\tau_T}(t_i) \gamma^{w_i} e^{-\gamma (r_i \wedge T - t_i)},
	\end{aligned}
\end{equation}
where \(T\) is the time horizon, and \(w_i\) is the binary censoring indicator as defined in \eqref{eq:lkh1}.

For the remaining \(n_1 + n_2\) partially incomplete records, the DSA likelihood function is based on \eqref{eq:f} and \eqref{eq:g}:
\begin{equation}\label{eq:lkh_inc}
\begin{aligned}
		&\mathcal{L}_I(\theta\vert t_1,\ldots,t_{n_1}, r_1,\ldots,r_{n_2}, T) \\ 
		&= \prod_{i=1}^{n_1} f_{\tau_T}(t_i) \prod_{i=1}^{n_2} g_{\tau_T}(r_i),
\end{aligned}
\end{equation}
where we assume \(r_i < T\). 

The overall likelihood for all \(n_T\) individual histories is obtained by multiplying \eqref{eq:lkh_comp} and \eqref{eq:lkh_inc}. Note that the likelihood expressions depend on the parameter \(\beta\) only implicitly, through the values of the survival function \(S(t)\) defined in \eqref{eq:dsa3}. 

The estimation of the parameter vector \(\theta\), as defined in \eqref{eq:theta}, involves maximizing the product of the two likelihood functions \eqref{eq:lkh_comp} and \eqref{eq:lkh_inc}. To enhance numerical stability, it is often advantageous to consider the formulation in terms of the logarithmic transformation \eqref{eq:logdsa}, which involves maximizing the sum of the logarithms of the two likelihood functions instead. This approach seamlessly integrates into the Bayesian estimation framework, allowing for more comprehensive propagation of uncertainty and the incorporation of external information into the statistical model. Consequently, this method yields parameter estimates that fully leverage all available information while accounting for uncertainty in a rigorous manner.

In our analysis of DRC data, the approximate posterior densities of \(\theta\) were obtained using the Hamiltonian Monte Carlo (HMC) sampler~\cite{monnahan2017faster}, implemented in the open-source statistical software STAN~\cite{carpenter2017stan}. This implementation was accessed via the R library Rstan~\cite{annis2017bayesian}. Following \cite{Vossler}, we assumed uniform (sometimes improper) prior distributions for the components of \(\theta\), specified as follows:
\begin{equation}
	\begin{aligned}
		&\tilde\beta \in (0.1, \infty), \\
		&\tilde\gamma \in (0, \beta), \\
		&\rho \in (0, 0.01).
	\end{aligned}
\end{equation}

As in \cite{Vossler}, the lower bound on \(\tilde\beta\) was informed by empirical data, while the upper bound on \(\gamma\) was chosen to enforce the constraint \(\tR > 1\). 

\begin{table}[tb]
    \caption{{\bf Parameter Estimates.} Comparison of posterior estimates (means and 95\% credibility bounds) from the classical SIR and network SIR   models for the DRC Ebola dataset. The parameters between two models are matched according  to Proposition~\ref{prop:1}.}
    \centering\small
    \begin{tabular}{c l l}
        \hline
        Param & Classical & Network  \\ [0.5ex]
        %heading
        \hline
        $\tilde\beta$ &   0.235 (0.218, 0.253) & 0.229 
        (0.209, 0.259) \\
        $\tilde\gamma$ &   0.214 (0.199, 0.230) & 0.215 
        (0.197,  0.242) \\
        $\rho$ &  0.0067 (0.0055, 0.0081) & 0.0055 (0.0046,  0.0073) \\
        $\mu$ &--&  39.48 (7.93, 93.00)  \\
        $\tR$ &  1.098 (1.061,1.135 ) & 1.071 (1.034, 1.109)\\
        % \hline
        $\hat{K}_\infty$ &   3481.41  & 3773.37  \\  & (2877.416, 4155.878) &(3373.245, 4226.315)\\ [0.5ex]
        \hline
    \end{tabular} \label{tab:1}
   
\end{table}
The  MCMC analysis  was based on  two  chains of 3,000 iterations, with a burn-in period of 1,000 iterations. Convergence of the chains was assessed using Rubin's \(R\) statistic~\cite{annis2017bayesian}. The analysis produced approximate samples from the posterior distribution of the parameter vector \(\theta^\ast\) (see Figure~\ref{fig:traces} below). 
\subsection{Results}

\begin{figure*}
\centering{\includegraphics[scale=.5]{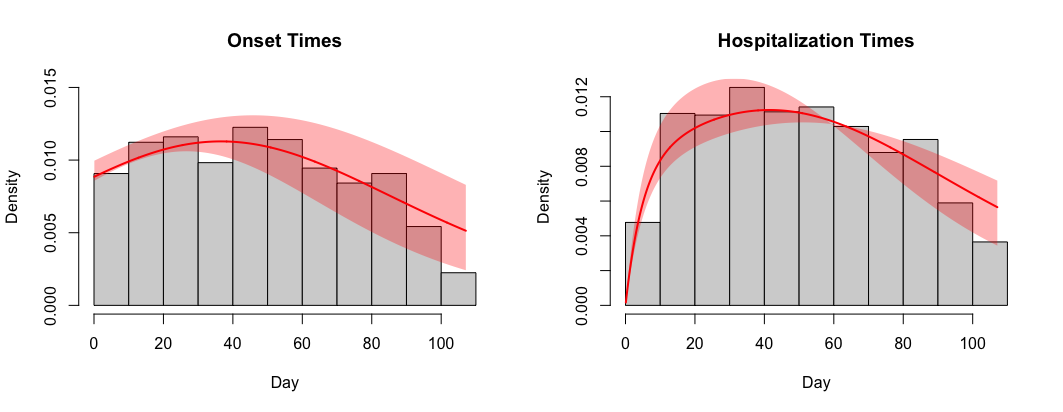}%
}
\caption{{\bf Model Fit.} Histograms of onset and recovery (hospitalization) times overlaid with Poisson SIR model fit, computed at the means of the posterior parameter values
(solid lines) with  the corresponding 95\% credible intervals  (shaded regions).}
\label{fig:fit}
\end{figure*}

The general comparison of the parametric predictions of the DSA model with the empirical data in the DRC data set between  May 27, 2019, and September
12, 2019 is given in Figure~\ref{fig:fit}, where the scaled theoretical densities of the epidemic are plotted alongside the observed relative daily counts of infection (onset) and removal (hospitalization). As seen from the plots, the network  model appears to capture well the  empirically  observed patterns of daily counts represented by  the histogram bars. The 95\% credible bounds  around the model fit (shaded in red)  are calculated based on the model parameter posterior distributions estimated via the MCMC algorithm with priors described in Section~\ref{ssec:par_est}. We note that, although  the network DSA model curve fit appears quite similar to that presented in Section~2 of \cite{Vossler}, the added benefit is the estimation of the distribution of an average degree  interpreted as an average number of contacts per infected individual prior to removal. Since the mean degree is estimated as relatively high the Proposition~\ref{prop:1} justifies also the use of classical SIR model  from \cite{Vossler} for the DSA analysis.

\begin{figure*}[h]	
	\centering
	\includegraphics[scale=.24]{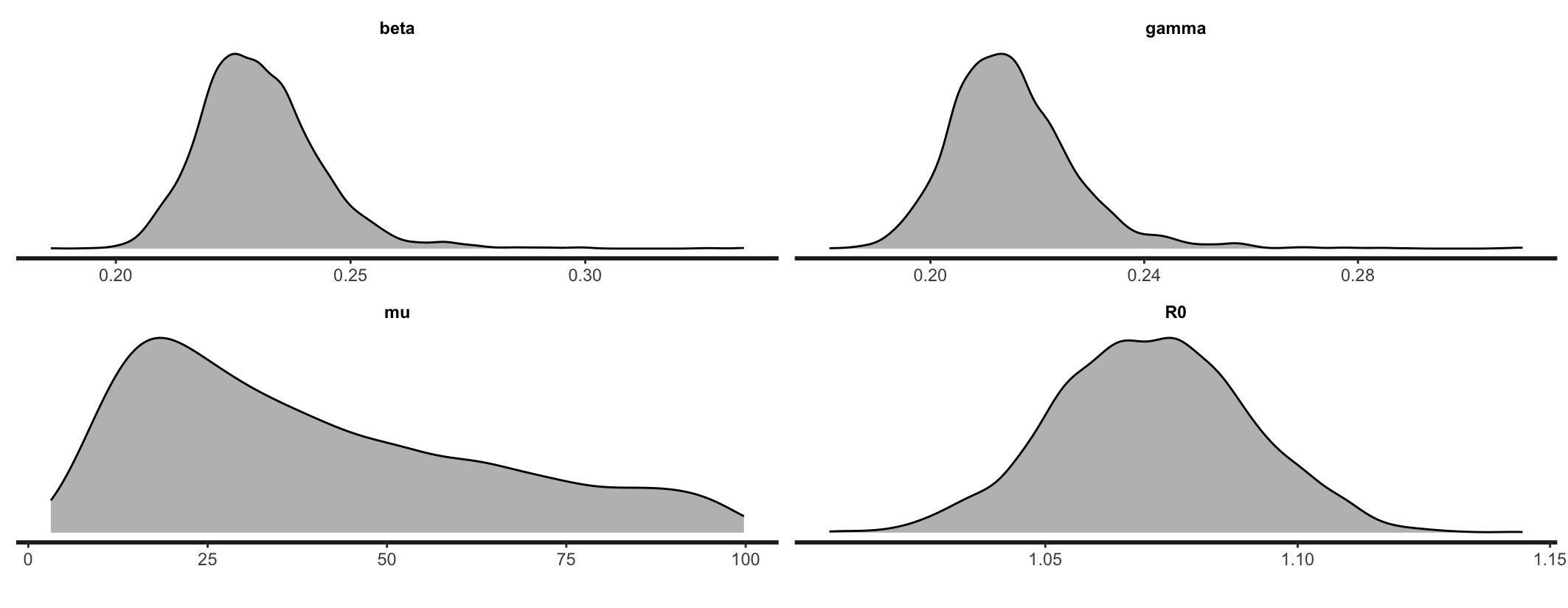}
	\caption{{\bf Posterior Parameter Densities.} The posterior distributions of $\tilde\beta$,  $\tilde\gamma$, $\mu$ and $\Rnet$. The posterior distribution of average degree ($\mu$) is seen to be right skewed but with mode below 25.}
  \label{fig:traces}
\end{figure*}

The  numerical  results of the MCMC analysis are summarized in  Table~\ref{tab:1} with some of the posterior plots presented in  Figure~\ref{fig:traces}. In Table~\ref{tab:1}, the posterior mean and corresponding credibility interval for each component of $\theta^\ast$ given by \eqref{eq:theta_pr} are listed  along with the estimated reproduction number (\(\tR\)). Additionally, in the last two rows, the posterior mean estimate of the outbreak size ($K_\infty$) as defined in \eqref{eq:khat} is  reported along with its 95\% credible bounds. The MCMC estimation scheme that produced the numerical values listed in the table  was based on  likelihood functions in equations~\eqref{eq:lkh_comp} and \eqref{eq:lkh_inc} conditioned on the observation periods  ($T=108$). 
Since the posterior estimate of the ratio $\tR/\mu\approx 1/40$ is relatively low,  the result of Proposition~\ref{prop:1} indicates that we should expect a reasonably good agreement between the classical mass-action SIR model \eqref{eq:sir2} and the Poisson network model \eqref{eq:dsa2}.  
Indeed, as seen from the entries of Table~\ref{tab:1}, the point parameter values (posterior means) for the infection rate $(\tilde\beta)$, recovery rate ($\tilde\gamma$) and the initial prevalence of infection $\rho$ all are numerically close for the classical and network models. In fact, since the respective credible intervals for $(\tilde\beta)$ and ($\tilde\gamma$) under network  model contain the credible intervals under classical SIR, one would conclude that the posterior estimates distributions are not statistically different. 
 
Although the   posterior means of the basic reproduction number  and  the final epidemic size are also close to the quantities obtained from the classical SIR model analysis, they posterior distributions are seen not to overlap and hence we conclude them to be different. 

   The final noteworthy result is that the SIR network model provides a better overall fit to the data, as indicated by the maximized likelihood function (not shown in the summary table). Additionally, the network analysis reveals that the posterior distribution of the average network degree ($\mu$) is highly right-skewed. This finding aligns with the observation that a small subset of infectious individuals had a disproportionately high number of contacts before their hospitalization.

Interestingly, the mean contact degree is approximately 40, while the posterior mode is slightly below 25 (see Figure~\ref{fig:traces}). For the range of parameters estimated from the data, the relatively high mean degree suggests that the two SIR models analyzed in Table~\ref{tab:1} are numerically quite similar in terms of the respective numbers of infected individuals. This similarity is further illustrated by the plots in Figure~\ref{fig:2} based on Proposition~1. 

\section{Discussion}\label{sec:4}

This paper presents a novel formulation of the classical SIR model, incorporating a dynamical survival analysis perspective into a configuration model based on a Poisson-distributed degree network. Among the broad class of configuration model networks, the Poisson network is particularly noteworthy due to its exact pair-level closure property. This characteristic makes it a representative model for many "appropriately regular" random networks (for instance, certain Bernoulli graphs) which often exhibit structural similarities to Poisson networks.

From a numerical perspective, the Poisson exact closure property prevents network complexity from scaling uncontrollably as the network size increases. This property enhances both analytical tractability and computational efficiency, offering a scalable and reliable approach for studying and managing disease dynamics. The proposed framework effectively models the contact patterns within a population that drive the spread of infectious diseases, providing deeper insights into outbreak mechanisms, disease persistence, and shifts in infection dynamics as interactions between infectious and susceptible individuals evolve.

The proposed model is broadly applicable to various domains, including social interactions, biological systems (e.g., neural or protein interactions), and technological networks (e.g., the spread of computer viruses or resilience of infrastructure systems). By transforming the SIR model using dynamical survival analysis within the edge-based configuration network framework, the resulting system of equations captures the intricate dynamics of network-based interactions. Despite the complexity of these interactions, the equations remain mathematically tractable, often enabling precise predictions of disease trends (see, for instance, the discussions of related  DSA-based approaches given in  \cite{RempalaDSA2023,Khud-COVID}).

The utility of the Poisson SIR network model is demonstrated through secondary analysis of the data from 2018–2020 Ebola outbreak in the Democratic Republic of the Congo. The model not only maintains a strong fit to empirical data but also reveals hidden structural features of the contact network underlying the disease's spread. Identifying such networks is crucial for effectively targeting at-risk populations—such as through vaccination campaigns—to prevent further transmission. This stochastic SIR framework thus provides a versatile tool for modeling infectious diseases and other dynamic processes beyond the scope of traditional SIR models.

Additionally, the study addresses an open challenge in approximating network-based SIR models with classical mass-action SIR models. Our empirical findings, encapsulated also in Proposition 1 reformulated after \cite{RempalaBiomath}, demonstrate that such an approximation is achievable by establishing a correspondence between the parameters of the two frameworks. Furthermore, it highlights that in networks of moderate to high degree, the classical SIR model can serve as a reasonable approximation for the susceptible and infected curves and a possible trend prediction tool.

Overall, our results  underscore the potential for bridging complex network-based dynamics with classical modeling approaches, offering both theoretical insights and practical tools for understanding and forecasting epidemic spread.

\subsection*{Author  Contributions} GAR and JW developed the model, analyzed the data, and drafted the initial manuscript. AG contributed to data analysis, manuscript revisions, and the development of the necessary software scripts. All authors reviewed and approved the final version of the manuscript. 

\subsection*{Data Availability} Deidentified dataset  
%and the code used in MCMC analysis are  publicly available via  Zenodo platform at XXXX %\url{https://doi.org/10.5281/zenodo.6104188}.
% \url{https://doi.org/10.5281/zenodo.5648029}.  
%Further data
 may be available upon a  reasonable request to the authors. 

\subsection*{Acknowledgement} The  work of GR and JW was partially funded by the grant from the HEALMOD initiative  at The Ohio State University. JW was partially funded by IMU-CDC.

\section*{Conflicts of Interest} The authors declare no conflicts of interest.  

%\newpage
%\bibliographystyle{unsrt}
%\bibliography{JRNets}

\end{document}